\documentclass[aps,pra,reprint,superscriptaddress]{revtex4-2}

\usepackage{graphicx}
\usepackage{amsfonts}
\usepackage{comment}
\usepackage{amsmath}
\usepackage{amsthm}

\usepackage[percent]{overpic} 

\usepackage{bm}
\usepackage{float}
\usepackage{hyperref}
\usepackage{nameref}
\usepackage{quantikz}
\usepackage[caption=false]{subfig}

\usepackage{xcolor}
\usepackage{soul}
\sethlcolor{yellow}
\soulregister\cite7
\soulregister\ref7
\soulregister\eqref7

\begin{document}

\title{Quasi-Monte Carlo Method for Linear Combination Unitaries via Classical Post-Processing}

\author{Yuya Kawamata}
\email[Email: ]{u597035b@ecs.osaka-u.ac.jp}
\affiliation{Graduate School of Engineering Science, The University of Osaka,
1-3 Machikaneyama, Toyonaka, Osaka 560-8531, Japan}
\author{Kosuke Mitarai}
\affiliation{Graduate School of Engineering Science, The University of Osaka,
1-3 Machikaneyama, Toyonaka, Osaka 560-8531, Japan}
\affiliation{Center for Quantum Information and Quantum Biology,
The University of Osaka, 1-2 Machikaneyama, Toyonaka 560-0043, Japan}
\email[Email: ]{mitarai.kosuke.es@osaka-u.ac.jp}
\author{Keisuke Fujii}
\affiliation{Graduate School of Engineering Science, The University of Osaka,
1-3 Machikaneyama, Toyonaka, Osaka 560-8531, Japan}
\affiliation{Center for Quantum Information and Quantum Biology,
The University of Osaka, 1-2 Machikaneyama, Toyonaka 560-0043, Japan}
\affiliation{RIKEN Center for Quantum Computing (RQC), Hirosawa 2-1, Wako, Saitama 351-0198, Japan}
\email[Email: ]{fujii.keisuke.es@osaka-u.ac.jp}

\date{\today} 

\begin{abstract}
We propose the quasi-Monte Carlo method for linear combination of unitaries via classical post-processing (LCU-CPP) on quantum applications. 
The LCU-CPP framework has been proposed as an approach to reduce hardware resources, expressing a general target operator $F(A)$ as $F(A) = \int_V f(\bm{t})\, G(A, \bm{t})\, d\bm{t}$, where each $G(A, \bm{t})$ is proportional to a unitary operator.
On a quantum device, $\mathrm{Re}[\mathrm{Tr}(G(A, \bm{t})\rho)]$ can be estimated using the Hadamard test and then combined through classical integration, allowing for the realization of nonunitary functions with reduced circuit depth.  
While previous studies have employed the Monte Carlo method or the trapezoid rule to evaluate the integral in LCU-CPP, we show that the quasi-Monte Carlo method can achieve even lower errors.
In two numerical experiments, ground state property estimation and Green’s function estimation, the quasi-Monte Carlo method achieves the lowest errors with a number of Hadamard test shots per unitary that is practical for real hardware implementations.  
These results indicate that quasi-Monte Carlo is an effective integration strategy within the LCU-CPP framework.
\end{abstract}

\maketitle

\section{INTRODUCTION}
Quantum computers hold the promise of surpassing classical computers in terms of computational power, offering revolutionary advances in fields such as quantum chemistry \cite{mcardle2020quantum, cao2019quantum}, financial engineering \cite{orus2019quantum, egger2020quantum}, and combinatorial optimization \cite{farhi2014quantum}.
Nevertheless, most algorithms with provable quantum speedups, such as the quantum linear system solver \cite{HHL} or Hamiltonian simulation \cite{georgescu2014quantum}, require fault-tolerant quantum computing (FTQC) and extensive error correction, which in turn demand substantial hardware resources \cite{babbush2018encoding, lee2021even, beverland2022assessing}.
To bridge the gap between near-term devices and these resource heavy algorithms, it is therefore critical to identify techniques that significantly reduce quantum hardware requirements \cite{preskill2018quantum}.

One promising approach is to offload as much work as possible to classical post-processing.
In particular, the “linear combination of unitaries via classical post-processing” (LCU-CPP) framework \cite{universal, universal2, inv2, gibbs1, gibbs2, diag,  Kyriienko2020Quantum, he2022quantum, keen2021quantumalgorithmsgroundstatepreparation, chakraborty2024implementing} lets us estimate quantities of the form $\mathrm{Tr}\bigl[F(A)\,\rho\bigr]$,
where $\rho$ is an $N$‐qubit quantum state and $A$ is a Hermitian matrix (e.g.\ a Hamiltonian), even when $F(A)$ itself is nonunitary. Concretely, one selects an integral representation $F(A) \;=\; \int_V f(\bm{t})\,G(A,\bm{t})\,d\bm{t}$, where $\bm{t}\in\mathbb{R}^d$, $V$ is a $d$‐dimensional hyperrectangular domain, $f(\bm{t})$ is a probability density function over $V$, and each $G(A,\bm{t})$ is a unitary.
Because $\mathrm{Tr}[F(A)\rho]$ is real, one can write $\mathrm{Tr}\bigl[F(A)\rho\bigr]= \int_Vf(\bm{t})\,\mathrm{Re}[\mathrm{Tr}\bigl(G(A,\bm{t})\,\rho\bigr)]d\bm{t}.$
On a quantum device, each term $\mathrm{Re}\bigl[\mathrm{Tr}(G(A,\bm{t})\,\rho)\bigr]$ is estimated by a Hadamard test.
These estimates are then combined by classical numerical integration.
In this way, complicated nonunitary functions, such as $A^{-1}$ for matrix inversion \cite{inv2, Kyriienko2020Quantum, keen2021quantumalgorithmsgroundstatepreparation}, $e^{-\beta A}$ for Gibbs state sampling \cite{universal2}, or filtered ground state projectors for property estimation \cite{universal, inv2, he2022quantum}, can be realized with reduced quantum circuit depth.
Despite this flexibility, prior LCU‐CPP studies simply invoke either the Monte Carlo (MC) method or the trapezoid rule for the integration, without comparing these options in detail.
However, the overall accuracy and resource cost of LCU-CPP depend sensitively on how one carries out the integral over $\bm{t}$.

In this work, we present the quasi-Monte Carlo (QMC) method \cite{qmcbook} for LCU-CPP.
More specifically, we first show the upper bounds for the errors under QMC, MC, and the trapezoid rule.  
When the number of Hadamard test shots per unitary is $M$, the analysis indicates that MC and QMC achieve the lowest error when $M=1$, QMC is most advantageous when $M$ is in a moderate range, and the trapezoid rule becomes favorable in the limit as $M \to \infty$.
To clarify what range of $M$ is practically moderate, we carry out numerical experiments on ground state property estimation and Green’s function estimation.
The results show that QMC attains the smallest error around $M=10^3$ in the former and around $M=10^2$ in the latter.
These shot counts can be considered reasonable for current hardware once execution and communication latencies are taken into account \cite{menickelly2022latency, ito2023latency}.  
In conclusion, QMC provides an effective integration strategy within the LCU-CPP framework.

The rest of the paper is organized as follows.
In Sec.~\ref{seq:pre}, we review the LCU-CPP framework and the error analysis of MC and the trapezoid rule.
In Sec.~\ref{seq:fa_qmc}, we introduce LCU-CPP via QMC and provide its advantages.
Sec.~\ref{sec:ne} presents numerical experiments on ground state properties and Green’s function, highlighting the practical gains of QMC.
Finally, Sec.~\ref{seq:con} summarizes our conclusion.

\section{Preliminaries}
\label{seq:pre}
In this section, we briefly review prior studies as preliminaries.
We first explain the framework of LCU-CPP, and then error estimates for LCU-CPP when using MC and the trapezoid rule, respectively.

\subsection{Linear combination of unitaries via classical post-processing (LCU-CPP)}
\label{subseq:pre_lcucpp}
\begin{figure}
\centering
    \centering
    \begin{quantikz}
        \lstick{$\ket{0}$} & \gate{H}\qw & \ctrl{1} & \gate{H}\qw & \meter{} \\
        \lstick{$\rho$} & \qwbundle[]{} & \gate{G(A,\bm{t})} & \qw & \qw
    \end{quantikz}
\caption{Quantum circuit for the Hadamard test. The measurement outcome $y\in{+1,-1}$ satisfies $\Pr[y=+1]=(1+x)/2$ and $\Pr[y=-1]=(1-x)/2$, where $x=\mathrm{Re}[\mathrm{Tr}(G(A,\bm t)\rho)]$. For more details, please refer to Appendix \ref{ap:hd}.}
\label{fig:hadamard_test}
\end{figure}
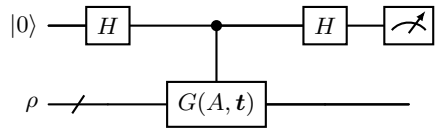

We consider the problem of estimating $\mathrm{Tr}(F(A)\rho)$ for an $N$-qubit quantum state $\rho$, a Hermitian matrix $A \in \mathbb{C}^{2^N\times 2^N}$ such as a Hamiltonian, and a general function $F: \mathbb{C}^{2^N\times 2^N} \to \mathbb{C}^{2^N\times 2^N}$.
To tackle this problem, the LCU-CPP framework relies on the following integral representation:
\begin{align}
F(A)  &= \int_{V} f(\bm{t}) G(A, \bm{t}) d\bm{t},
\label{eq:basic}
\end{align}
where $\bm{t} \in \mathbb{R}^{d}$, $V$ is a $d$-dimensional hyperrectangular domain $V = \prod_{j=1}^{d}[a_j, b_j]$ of volume $|V| = \prod_{j=1}^d (b_j - a_j)$, $f(\bm{t})$ is a probability density function, and $G(A, \bm{t})$ is a constant multiplied unitary operator parameterized by $A$ and $\bm{t}$.
While this integral expansion is not universally applicable for every possible $F(A)$, many important cases do admit such a form, as we see below.
From Eq.~\eqref{eq:basic}, it follows that
\begin{align}
\mathrm{Tr}\left(F(A)\rho\right)=\int_{V}f(\bm{t})\,\mathrm{Tr}\left(G(A, \bm{t})\rho\right)\, d\bm{t}.   
\label{eq:basic2}
\end{align}
Since $F(A)$ is assumed to be Hermitian, $\mathrm{Tr}(F(A)\rho)$ is real, which allows us to write $\mathrm{Tr}(F(A)\rho) = \int_{V} f(\bm{t})\mathrm{Re}[\mathrm{Tr}(G(A, \bm{t})\rho)]d\bm{t}$.
Because $\mathrm{Re}[\mathrm{Tr}(G(A, \bm{t})\rho)]$ can be estimated via the Hadamard test (see Fig.~\ref{fig:hadamard_test}), we can estimate $\mathrm{Tr}(F(A)\rho)$ by numerical integration over $V$.
We refer to this procedure as LCU-CPP.
This idea appears in a number of studies, dating back at least to 1999 \cite{deraedt1999quantum}.
It is worth noting that one could also implement these integrals by performing linear combinations of unitaries entirely on a quantum computer, as in \cite{Childs2017Quantum}; however, using the Hadamard test together with classical post-processing typically reduces the hardware resources.
Below, we briefly highlight three representative applications of this approach.

\textbf{Microcanonical ensemble ($d=1$).}
First, we consider estimating the expectation value of a unitary $O$ in a microcanonical ensemble at energy $E$ using LCU-CPP \cite{gibbs1, gibbs2}, which involves calculating $\mathrm{Tr}(Oe^{-(A-E)^2 \tau^2}\rho)$ with $\rho = I/2^N$.
As in Eq.~\eqref{eq:basic2}, we use
\begin{align}
F(A) &= Oe^{-(A-E)^2 \tau^2} \\
f(t) &= \frac{1}{2\tau\sqrt{\pi}} e^{-\frac{t^2}{4\tau^2}} \\
G(A,t) &= O\,e^{-i(A-E)t}, 
\end{align}
where $\tau$ is a constant.
The integration domain is an infinite domain, which is reduced to a finite region by introducing an appropriate cutoff.

\textbf{Ground state property estimation ($d=2$).}
Next, we consider estimating the expectation value of a unitary $O$ on the ground state $\rho_{g}$ using LCU-CPP \cite{universal, inv2, he2022quantum}, which involves calculating $\mathrm{Tr}(e^{-A^2 \tau^2}\, O e^{-A^2 \tau^2} \rho)$. 
We prepare an unnormalized ground state by applying a Gaussian function to a state $\rho$ that has nonzero overlap with the ground state. 
As in Eq.~\eqref{eq:basic2}, we use
\begin{align}
F(A) &= e^{-A^2 \tau^2}\, O \, e^{-A^2 \tau^2}, \\
f(\bm t) &= \frac{1}{4\tau^2 \pi} \, e^{-\tfrac{t_1^2+{t_2}^2}{4 \tau^2}}, \\
G(A,\bm t) &= e^{-iAt_1}\,O \,e^{-iA t_2},
\end{align}
where $\bm t=(t_1,t_2)$.
The integration domain is an infinite domain in two dimensions, which is reduced to a finite region by introducing appropriate cutoffs.

\textbf{Linear system solver ($d=4$).}
Lastly, we consider estimating $\mathrm{Tr}(A^{-1} O A^{-1}\rho)$ using LCU-CPP for an invertible Hermitian matrix $A$ and a unitary $O$ \cite{inv2}. 
Here, $A$ is rescaled so that $\mathrm{spec}(A)\subset[-1,-\mu]\cup[\mu,1]$ with $\mu>0$.
The integration domain is $V=[0,L_{1}]\times[-L_{2},L_{2}]\times[0,L_{3}]\times[-L_{4},L_{4}]$, and we define $Z(L):=1-e^{-L^{2}/2}$ and $\mathrm{sign}(x)$ as the sign function.
With normalization chosen so that $f$ is a probability density after truncation, we set, as in Eq.~\eqref{eq:basic2},
\begin{align}
F(A) &= A^{-1} O A^{-1}, \\
f(\bm t)
&=\frac{1}{4L_{1}L_{3}Z(L_{2})Z(L_{4})}\,|t_{2}t_{4}|\,e^{-\frac{t_{2}^{2}+t_{4}^{2}}{2}}, \\
G(A,\bm t)
&=-\frac{2Z(L_{2})Z(L_{4})}{\pi} \notag \\
&\ \ \ \ \times \mathrm{sign}(t_{2}t_{4})\,L_{1}L_{3}\,e^{-i t_{1} t_{2} A}\,O\,e^{-i t_{3} t_{4} A},
\end{align}
where $\bm t=(t_1,t_2,t_3,t_4)$.
This representation recovers the untruncated one as $L_j\to\infty$.

\subsection{LCU-CPP via Monte Carlo Method (MC)}
\label{sec:mc}
MC is widely used for numerical integration and has been employed in several works \cite{universal,universal2, gibbs1, diag}.
Within this framework, one can estimate $\mathrm{Tr}(F(A)\rho)$ by sampling random $\bm{t}$ according to the function $f(\bm{t})$.
More concretely, we first draw $K$ samples $\bm{t}_{k}^{\mathrm{mc}}$ from the distribution proportional to $f(\bm{t})$.
On each $\bm{t}_{k}^{\mathrm{mc}}$, we perform the Hadamard test $M$ times to obtain estimators $\nu_k^{\mathrm{mc}}$ that satisfy $|\nu_k^{\mathrm{mc}} - \mathrm{Re}[\mathrm{Tr}(G(A, \bm{t}_k^{\mathrm{mc}})\rho)]| \leq \mathcal{O} (1/\sqrt{M})$ with high probability.
Finally, we form the estimator $\nu^{\mathrm{mc}} = \frac{1}{K}\sum_{k=1}^{K}\nu_k^{\mathrm{mc}}$, which serves as an approximation to $\mathrm{Tr}(F(A)\rho)$.
In Appendix~\ref{ap:mc}, we prove that
\begin{align}
\mathrm{Ex}[\nu^{\rm mc}] &= \mathrm{Tr}(F(A)\rho),\\
\mathrm{Var}[\nu^{\rm mc}] &= \frac{C^{\rm mc, ht}}{MK} + \frac{C^{\rm mc, ni}}{K},
\end{align}
where
\begin{align}
C^{\rm mc, ht} &= 1-\int_{V} f(\bm{t}) \mathrm{Re}[\mathrm{Tr}(G(A,\bm{t})\rho)]^2 d\bm{t}, \\
C^{\rm mc, ni} &= \int_V f(\bm{t})\mathrm{Re}[\mathrm{Tr}(G(A,\bm{t})\rho)]^2 d\bm{t} - \mathrm{Tr}(F(A)\rho)^2.
\end{align}
Here, the term $C^{\mathrm{mc,ht}}/MK$ represents the variance contribution due to the Hadamard test, and $C^{\mathrm{mc,ni}}/K$ represents the variance contribution due to numerical integration.
Consequently, the former scales as $\mathcal{O}(1/(MK))$, and the corresponding statistical error scales as $\mathcal{O}(1/M^{1/2}K^{1/2})$, while the latter scales as $\mathcal{O}(1/K)$, and the corresponding numerical integration error scales as $\mathcal{O}(1/K^{1/2})$.

\subsection{LCU-CPP via the Trapezoid Rule}
\label{sec:tr}
The trapezoid rule is another common approach to numerical integration and has been employed in several studies \cite{inv2, gibbs2, Kyriienko2020Quantum, he2022quantum, keen2021quantumalgorithmsgroundstatepreparation}.
According to this method, one uses a set of deterministic grid points $\{\bm{t}_k^{\mathrm{tr}}\}$ over the domain $V$.
Concretely, we take $K=\prod_{j=1}^d K_j$ grid points, where $K_j$ is the number of integration points along the $j$th dimension. 
For each integration point $\bm{t}_{k}^{\mathrm{tr}}$, we perform the Hadamard test $M$ times to obtain estimators $\nu_{k}^{\mathrm{tr}}$ satisfying $|\nu_k^{\rm{tr}} - \mathrm{Re}[\mathrm{Tr} (G(A, \bm{t}_k^{\rm tr})\rho)]| < \mathcal{O}(1/\sqrt{M})$.
Let $\omega_{k}$ be the trapezoid weight for each $\bm{t}_{k}^{\mathrm{tr}}$. Then our final estimator is $
\nu^{\rm tr} = \prod_{j=1}^{d} (b_j - a_j)/(K_j - 1)\sum_{k=1}^{K} \omega_k f(\bm{t}_k^{\rm tr}) \nu_k^{\rm{tr}}$, which approximates $\mathrm{Tr}(F(A)\rho)$.
In Appendix~\ref{ap:tr}, we show
\begin{align}
\mathrm{Ex}[\nu^{\rm tr}] = \mathrm{Tr}(F(A)\rho) + \mathrm{Bias}^{\rm{tr}}(K),
\end{align}
where $\mathrm{Bias}^{\mathrm{tr}}(K)$ arises from the numerical integration. 
Hence, unlike the Monte Carlo estimator, this method is biased, and we must consider the mean square error (MSE) rather than just the variance:
\begin{align}
\mathrm{MSE}^{\rm tr} &= \mathrm{Ex}[\left(\nu^{\rm tr} - \mathrm{Tr}[F(A)\rho] \right)^2] \notag \\
\label{eq:mse_tr}
&= \mathrm{Var}[\nu^{\rm tr}] + \mathrm{Bias}^{\rm tr}(K)^2,
\end{align}
where
\begin{align}
\label{eq:tr_var}
\mathrm{Var}[\nu^{\rm tr}] &\leq \frac{C^{\rm tr, ht}}{MK}+ \mathcal{O}\left(\frac{1}{MK^{\frac{2}{d}+1}}\right), \\
C^{\rm tr, ht} &= 2^d |V|\int_{V} f^2(\bm{t}) \left(1 - \mathrm{Re}[\mathrm{Tr}(G(A,\bm{t})\rho)]^2\right) d\bm{t},
\label{eq:tr_hd}
\end{align}
and
\begin{align}
\label{eq:tr_bias}
&|\mathrm{Bias}^{\rm tr}(K)| \leq \frac{C^{\rm tr, ni} }{K^{\frac{2}{d}}} + \mathcal{O}\left(\frac{1}{K^{\frac{3}{d}}}\right), \\
&C^{\rm tr, ni} = \sum_{j, j'=1}^d 
\frac{|V|}{3} \max_{\bm{t}} \Bigl|\frac{\partial^2}{\partial t_j \partial t_{j'}} f(\bm{t}) \mathrm{Re}\bigl[\mathrm{Tr}\bigl(G(A,\bm{t})\,\rho\bigr)\bigr]\Bigr| \notag  \\ &\ \ \ \ \ \ \ \ \ \ \ \times (b_j - a_j)\,(b_{j'} - a_{j'}).
\end{align}
Here, $\mathrm{Var}[\nu^{\mathrm{tr}}]$ represents the variance due to the Hadamard test, and $\mathrm{Bias}^{\rm tr}(K)$ represents the numerical integration error.
Consequently, the variance scales as $\mathcal{O}(1/MK)$, and the corresponding statistical error scales as $\mathcal{O}(1/M^{1/2}K^{1/2})$, while the numerical integration error scales as $\mathcal{O}(1/K^{2/d})$.

\section{LCU-CPP VIA QUASI-MONTE CARLO METHOD}
\label{seq:fa_qmc}
In this section, we propose the LCU-CPP using QMC.
We begin with an overview of QMC in classical numerical integration, and then analyze the mean squared error of LCU-CPP when employing QMC.
Finally, we highlight that QMC can possibly achieve faster convergence in integration error compared to MC, as well as a smaller constant coefficient than that associated with the trapezoid rule.

\subsection{Quasi-Monte Carlo Method (QMC)}
First, we introduce QMC in classical numerical integration theory~\cite{qmcbook}.
This method improves the asymptotic error behavior of MC by using low-discrepancy sequences, such as the Halton sequence, instead of random sampling.
Let $h: \mathbb{R}^d \to \mathbb{R}$ be an integrand, and $p(\bm{x})$ a probability density function over a $d$-dimensional hyperrectangular domain $V$.
We aim to approximate the integral
\begin{align}
I = \int_V p(\bm{x}) h(\bm{x}) d\bm{x}    
\end{align}
by the QMC estimator
\begin{align}
Q(K) = \frac{1}{K}\sum_{k=1}^{K} h(\bm{u}_k),
\end{align}
where each $\bm{u}_k = \mathrm{CDF}_p^{-1}(\bm{x}_k)$ is generated by applying the inverse cumulative distribution function (CDF) of $p(\bm{x})$ to a low-discrepancy sequence $\{\bm{x}_k\}$ such as the Halton sequence on $[0,1]^d$.
Here, $\mathrm{CDF}_p^{-1}$ denotes the inverse of the multivariate CDF of $p$, which can be constructed component-wise if $p$ is a product distribution (i.e., if the variables are independent).
This procedure is analogous to inversion sampling in MC.
The Koksma-Hlawka inequality~\cite{qmcbook} gives the following bound:
\begin{align}
\left| I - Q(K) \right| \leq V_{\rm HK} (h) \times D^*(\{\bm{u}_k\}, K),
\end{align}
where $V_{\rm HK}(h)$ is the variation of $h$ in the sense of Hardy and Krause, and $D^*(\{\bm{u}_k\}, K)$ is the star-discrepancy of the point set $\{\bm{u}_k\}$.
More explicitly,
\begin{align}
V_{\rm HK}(h) &= \sum_{r} \int_{V_{r}} \left| \frac{\partial^{|\bm{r}|} h(\bm{x}_r, \bm{1})}{\partial \bm{x}_r} \right| d\bm{x}_r, \\
D^*(\{\bm{u}_k\}, K) &= \sup_{J' \subset V} \left| \frac{1}{K} \#\{i \leq K : \bm{u}_i \in J'\} - \mathrm{Vol}(J') \right|,
\end{align}
where $r$ runs over all nonempty subsets of $\{1,\dots,d\}$, and $V_r = \left\{ (x_1, \dots, x_d) \in V \mid x_j = 1 \text{ for all } j \notin r \right\}$.
In $h(\bm{x}_r, \bm{1})$, the coordinates not in $r$ are set to $1$.
The operator $\frac{\partial^{|r|}}{\partial\bm{x}_r}$ denotes the mixed partial derivative with respect to all variables in $r$.
The supremum is over all axis-aligned subrectangles $J' \subset V$ of the form $J' = \mathrm{CDF}_p^{-1}([0,\bm{t}])$ for $\bm{t} \in [0,1]^d$, and $\mathrm{Vol}(J')$ is the Lebesgue measure of $J'$.
Theorem~2.35 in Ref.~\cite{qmcbook} shows that the star discrepancy of the Halton sequence satisfies $D^*(\{\bm{x}_k\}, K) = \mathcal{O}(\log^d K / K)$.
In this paper, we assume that $D^*(\{\bm{u}_k\}, K) = \mathcal{O}(\log^d K / K)$ also holds for the transformed sequence $\{\bm{u}_k\}$.
Although this does not hold for arbitrary $p$, this can be numerically verified for each case before performing integration for $h(\bm{x})$.
Since $V_{\mathrm{HK}}(h)$ depends only on the integrand, the total integration error satisfies 
\begin{align}
|I-Q(K)|=\mathcal{O}\left(\frac{\log^{d}K}{K}\right),    
\end{align}
which is asymptotically faster than the MC rate.
As a remark, more tigher error bounds could be obtained by adopting the Reproducing Kernel Hilbert Space (RKHS) framework or by considering quasi-uniform sequences other than the Halton sequence.
Nevertheless, we employ the simplest error estimate in this work, which does not affect the generality of the arguments that follow.

\subsection{LCU-CPP via Quasi-Monte Carlo Method}
\label{sec:lcu_qmc}
We now describe the LCU-CPP framework using QMC. 
For LCU-CPP via QMC, the mean square error (MSE) to an estimator represents the total error. 
We show that this MSE consists of a variance arising from the statistical error and a bias term arising from the numerical integration error.
At first, the estimator is defined as
\begin{align}
\nu^{\rm qmc} = \frac{1}{K} \sum_{k=1}^{K} \nu_k^{\rm qmc},
\end{align}
where
\begin{align}
\nu_k^{\rm qmc} = \frac{1}{M} \sum_{m=1}^{M} s_{m}(\bm{t}_k^{\rm qmc}).
\end{align}
Here, $s_{m}(\bm{t}_k^{\rm qmc}) \in \{+1, -1\}$ is the outcome of the $m$th Hadamard test, which is used to estimate $\mathrm{Re}\left[\mathrm{Tr}\bigl(G(A, \bm{t}_k^{\rm qmc}) \rho \bigr)\right]$.  
A low-discrepancy sequence $\{\bm{t}_k^{\rm qmc}\}$ is generated by $\bm{t}_k^{\rm qmc} = \mathrm{CDF}_p^{-1}(\bm{x}_k)$, where $\mathrm{CDF}_p(\bm{x}) = \int_{-\infty}^{x} p(\bm{x}')\, d\bm{x}'$, and $\{\bm{x}_k\}$ is the Halton sequence in $[0,1]^d$.

Next, we derive the mean square error of $\nu^{\rm qmc}$ with respect to $\mathrm{Tr}[F(A)\rho]$.  
First, note that for each fixed $\bm{t}_k^{\rm qmc}$, it holds that $\mathrm{Ex}[s_{m}(\bm{t}_k^{\rm qmc})] = \mathrm{Re}[\mathrm{Tr}(G(A, \bm{t}_k^{\rm qmc}) \rho)]$.  
Therefore, 
\begin{align}
\mathrm{Ex}[\nu_k^{\rm qmc}] &= \frac{1}{M} \sum_{m=1}^{M} \mathrm{Ex}[s_{m}(\bm{t}_k^{\rm qmc})] \\
&= \mathrm{Re}[\mathrm{Tr}(G(A, \bm{t}_k^{\rm qmc}) \rho)],
\end{align}
and so
\begin{align}
\mathrm{Ex}[\nu^{\rm qmc}] &= \frac{1}{K} \sum_{k=1}^{K} \mathrm{Ex}[\nu_k^{\rm qmc}] \\
&= \frac{1}{K} \sum_{k=1}^{K} \mathrm{Re}[\mathrm{Tr}(G(A, \bm{t}_k^{\rm qmc}) \rho)].    
\end{align}

Let us define the numerical integration error as $\mathrm{Bias}^{\rm qmc}(K) := \frac{1}{K} \sum_{k=1}^{K} \mathrm{Re}[\mathrm{Tr}(G(A, \bm{t}_k^{\rm qmc}) \rho)] - \int_V f(\bm{t}) \mathrm{Re}[\mathrm{Tr}(G(A, \bm{t}) \rho)]\, d\bm{t} = \mathrm{Ex}[\nu^{\rm qmc}] - \mathrm{Tr}[F(A)\rho]$, so that
\begin{align}
\mathrm{Ex}[\nu^{\rm qmc}] = \mathrm{Tr}[F(A)\rho] + \mathrm{Bias}^{\rm qmc}(K).
\end{align}

Thus, as with the trapezoid rule, this estimator is biased, and we consider the mean square error (MSE) rather than just the variance.  
According to the Koksma–Hlawka inequality, we have $|\mathrm{Bias}^{\rm qmc}(K)| \leq C^{\rm qmc, ni} \frac{(\log K)^d}{K}$,  where $C^{\rm qmc, ni}$ is a constant involving $V_{\mathrm{HK}}(\mathrm{Re}[\mathrm{Tr}(G(A, \bm{t})\rho)])$.

Now, let us consider the variance of $\nu^{\rm qmc}$.  
Since $\mathrm{Var}[s_{m}(\bm{t}_k^{\rm qmc})] = \mathrm{Ex}[s_{m}(\bm{t}_k^{\rm qmc})^2] - \left(\mathrm{Ex}[s_{m}(\bm{t}_k^{\rm qmc})]\right)^2 = 1 - \mathrm{Re}[\mathrm{Tr}(G(A, \bm{t}_k^{\rm qmc}) \rho)]^2$, and using $\int_V f(\bm{t})\left(1 - \mathrm{Re}[\mathrm{Tr}(G(A, \bm{t})\rho)]^2\right) d\bm{t} = \frac{1}{K} \sum_{k=1}^{K} \left(1 - \mathrm{Re}[\mathrm{Tr}(G(A, \bm{t}_k^{\rm qmc})\rho)]^2\right) + \mathcal{O}((\log K)^d / K)$,
we find
\begin{align}
\mathrm{Var}[\nu^{\rm qmc}] &= \mathrm{Var}\left[\frac{1}{K} \sum_{k=1}^{K} \nu_k^{\rm qmc}\right] \\
&= \frac{1}{K^2 M^2} \sum_{k=1}^{K} \sum_{m=1}^{M} \mathrm{Var}[s_{m}(\bm{t}_k^{\rm qmc})] \\
&= \frac{1}{K^2 M} \sum_{k=1}^{K} \left(1 - \mathrm{Re}[\mathrm{Tr}(G(A, \bm{t}_k^{\rm qmc})\rho)]^2\right) \\
\label{eq:qmc_var}
&= \frac{1}{K M} \int_{V} f(\bm{t})\left(1 - \mathrm{Re}[\mathrm{Tr}(G(A, \bm{t})\rho)]^2\right) d\bm{t} \notag \\
&\qquad + \mathcal{O}\left(\frac{(\log K)^d}{K^2 M}\right).
\end{align}

Combining this variance with the bias, the MSE as the total error in LCU-CPP via QMC is given by
\begin{align}
\mathrm{MSE}^{\rm qmc} &= \mathrm{Ex}\left[\left(\nu^{\rm qmc} - \mathrm{Tr}[F(A)\rho] \right)^2\right] \\
\label{eq:mse_qmc}
&= \mathrm{Var}[\nu^{\rm qmc}] + \mathrm{Bias}^{\rm qmc}(K)^2.
\end{align}
Here, $\mathrm{Var}[\nu^{\rm qmc}]$ represents the variance due to the Hadamard test, while $\mathrm{Bias}^{\rm qmc}(K)$ represents the numerical integration error.
Consequently, the variance scales as $\mathcal{O}(1/MK)$, and the corresponding statistical error scales as $\mathcal{O}(1/M^{1/2}K^{1/2})$, while the numerical integration error scales as $\mathcal{O}((\log K)^d/K)$.

\subsection{Advantages}
\label{seq:qmc_ad}
Let us discuss the advantages of QMC.
The convergence rates for each method are summarized as follows:
\begin{itemize}
    \item \textbf{Monte Carlo} (see Sec.~\ref{sec:mc}): \\
    Statistical error: $\mathcal{O}(1/M^{1/2} K^{1/2})$ \\
    Numerical integration error: $\mathcal{O}(1/K^{1/2})$
    \item \textbf{Trapezoid Rule} (see Sec.~\ref{sec:tr}): \\
    Statistical error: $\mathcal{O}(1/M^{1/2} K^{1/2})$ \\
    Numerical integration error: $\mathcal{O}(1/K^{2/d})$
    \item \textbf{Quasi-Monte Carlo} (see Sec.~\ref{sec:lcu_qmc}): \\
    Statistical error: $\mathcal{O}(1/M^{1/2} K^{1/2})$ \\
    Numerical integration error: $\mathcal{O}((\log K)^d/K)$
\end{itemize}
Compared to MC, QMC achieves a numerical integration error of $\mathcal{O}((\log K)^d/K)$, which converges asymptotically faster than MC, although both methods have the same rate for the statistical error.
Therefore, when the number of Hadamard test shots $M$ is large and the statistical error is ignorable, the quasi-Monte Carlo method can outperform the Monte Carlo method.
Comparing with the trapezoid rule, the trapezoid rule achieves a numerical integration error of $\mathcal{O}(1/K^{2/d})$, which is almost faster or the same as QMC, especially for $d=2$ as in many LCU-CPP applications.
However, while the statistical errors for both QMC and the trapezoid rule converge as $\mathcal{O}(1/M^{1/2} K^{1/2})$, the constant factor for the trapezoid rule depends on the integration domain $|V|$, as shown in Eq.~(\ref{eq:tr_hd}).
This dependency can be particularly disadvantageous when integrating over large domains, which is often encountered in LCU-CPP applications.
In contrast, QMC can avoid this disadvantage by using importance sampling, which makes it independent of the integration domain.

As a result, this leads to the following scenarios:
\begin{itemize}
    \item \textbf{
    When $M = 1$:}
    \\
    The statistical error dominates. MC and QMC achieve the lowest error.
    \item \textbf{
    When $M$ is moderate (e.g., $10^2$ or $10^3$):}
    The statistical error and numerical integration error are balanced. QMC achieves the lowest error.
    \item \textbf{
    When $M \to \infty$:}
    \\
    The numerical integration error dominates. the trapezoidal rule achieves the lowest error.
\end{itemize}
In many previous studies, the number of Hadamard test shots is fixed to $M=1$.
In realistic experimental setups, however, $M$ should be regarded as a tunable parameter. Equivalently, the relevant resource is the total number of shots $MK$.
This viewpoint is natural when taking into account experimental constraints such as the latency associated with quantum circuit execution and information transmission~\cite{menickelly2022latency, ito2023latency}, under which repeated sampling is often less costly than circuit preparation.
Within this framework, QMC is expected to become advantageous when $M$ takes a moderate value.
At this point, the question then is what constitutes an appropriate choice of $M$.
Since both the statistical error and the numerical integration error depend on the integrand, it is necessary to consider concrete applications.
While the range of admissible values of $M$ is determined by the hardware experimental setup, it is nonetheless useful to clarify the values of $M$ for which QMC becomes practically advantageous.
In the next section, we perform numerical experiments on two representative applications, which substantiate this scenario and provide explicit examples of values of $M$ for which QMC is practically favorable within the framework.

It should also be noted that the numerical integration error of the trapezoid rule can be significantly reduced by using extended Newton-Cotes formulas, such as Simpson's rule~\cite{davis2007methods}.
These methods only require modifying the weights in the trapezoid rule, and can also be applied within the LCU-CPP framework, with a similar error analysis.
However, while Newton-Cotes rules can provide theoretical improvements in accuracy, they tend to become increasingly numerically unstable for higher-order formulas.
In our experiments, we also include results for Simpson's rule as a reference, as it is the most well-known of the Newton-Cotes formulas.

\section{NUMERICAL EXPERIMENT}
\label{sec:ne}
In this section, we conduct two numerical experiments, estimating ground state properties and the Green's function, to demonstrate that the quasi-Monte Carlo method yields the lowest error in LCU-CPP applications.
Our results show that the QMC achieves the smallest errors, owing to its lower coefficients for numerical integration error compared to the trapezoid and Simpson's rules, and its faster convergence relative to the MC in moderate values $M$.
All source code and data are publicly available on the Zenodo repository~\cite{zenodo_code}.

\subsection{Case 1: Ground State Properties}
We first present a numerical experiment aimed at estimating properties of the ground state in a quantum many-body system \cite{universal, inv2, he2022quantum}.
The key idea is to apply a filter function $e^{-\tau^{2} H^{2}}$ to a quantum state.
Since this filter suppresses higher energy state more rapidly than the ground state, the ground state is selectively amplified.
Here, $H$ denotes the positive definite Hamiltonian of the system, and $\tau$ is a tunable parameter.
More precisely, let $\rho_0$ be the initial quantum state and let $\rho_g$ denote the ground state of $H$.
The expectation value of an observable $O$ with respect to the ground state, $\mathrm{Tr}[O\rho_{g}]$, can be approximated as $\mathrm{Tr}[ e^{-\tau^{2} H^{2}} O e^{-\tau^{2} H^{2}} \rho_0]/\mathrm{Tr}[ e^{-2\tau^{2} H^{2}} \rho_0 ]$.
In order to estimate the numerator and denominator respectively using the LCU-CPP, we use a Fourier representation of the Gaussian filter: $ \exp(-\tau^2H^2) = \int_{-\infty}^{\infty} 1/\sqrt{4\pi\tau^2} \exp(-t^2/4\tau^2) \exp(-iHt) dt$.
Then the numerator that we estimate here is expressed in the format of Eq. (\ref{eq:basic}) as:
\begin{align}
F(H) &= e^{-\tau^{2}H^{2}}Oe^{-\tau^{2}H^{2}}, \\
f(\bm{t}) &= \frac{1}{4\pi\tau^{2}}e^{-\frac{t^{2}+t'^{2}}{4\tau^{2}}}, \\
G(H,\bm{t}) &= e^{-iH t}Oe^{-iH t'}.
\end{align}
For our experiment, we adopt the one-dimensional Heisenberg model as the system Hamiltonian, and employ a two-site correlator as the observable:
\begin{align}
\label{eq:heisenberg}
H &= J \sum_{j=0}^{N-1} \left( X_j X_{j+1} + Y_j Y_{j+1} + Z_j Z_{j+1} \right) + E_0 I, \\
\rho &= \rho_0 = \frac{1}{2^{N}}\sum_{x=0}^{2^{N}-1}| x\rangle\langle x|, \\
O &= Z_0 Z_1,
\end{align}
where periodic boundary conditions are assumed: site $N$ is identified with site $0$.
The parameters used are $N=6$, $J=1$, and $\tau = 8$. 
The offset $E_0$ is set to the ground state energy, which is computed in advance.
For numerical integration using the trapezoid and Simpson's rules, we introduce a cutoff region $[-t_c, t_c] \times [-t'_c, t'_c]$, where the cutoff $t_c$ is chosen such that the absolute error is set to the double-precision floating-point accuracy, and we use $t_c = t'_c = 100$.
In the simulation of the Hadamard test, we do not consider the error from the Trotter decomposition or other hardware errors.

\begin{figure}[t]
    \centering
    \includegraphics[width=0.9\linewidth]{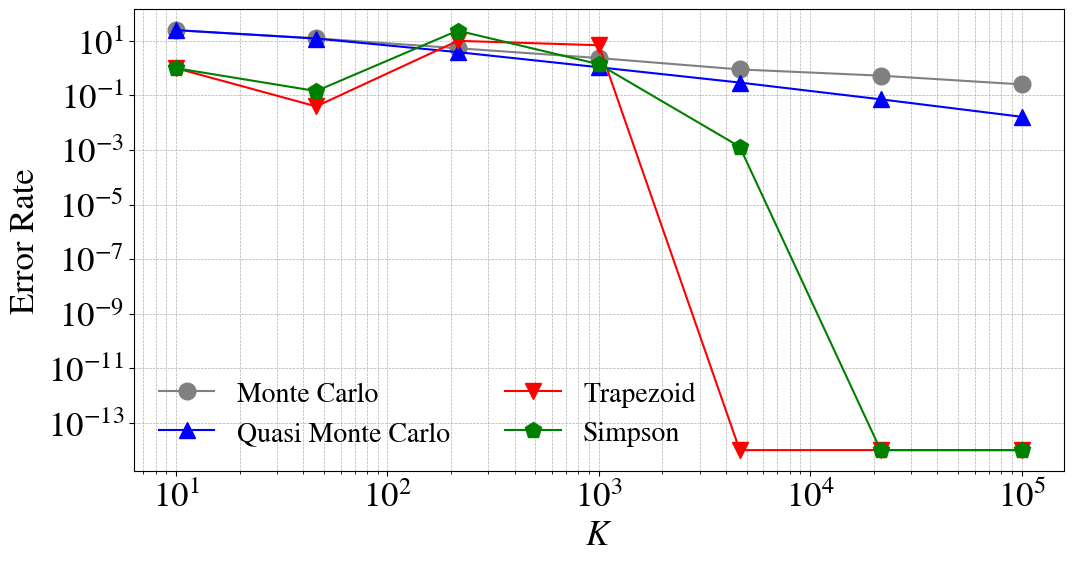}
    \caption{Error rate versus the number of the different Hadamard tests $K$ for each integration method without statistical error for estimating the ground state properties. This corresponds to the limit case where $M \rightarrow \infty$.}
    \label{fig:ground_exact}
\end{figure}

\begin{figure}[t]
    \centering
    \begin{overpic}[width=0.9\linewidth]{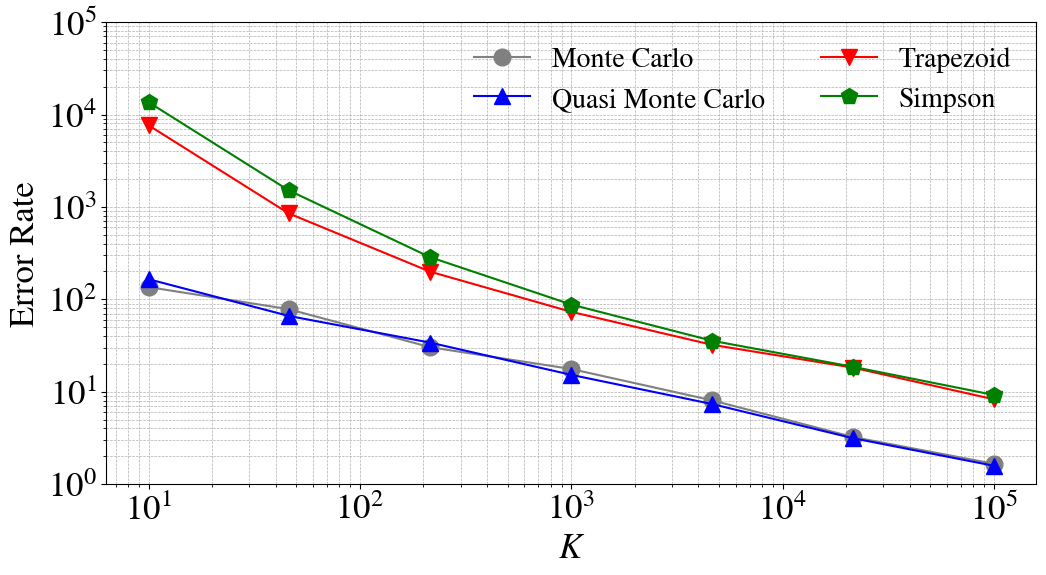}
        \put(1,55){\textbf{(a)}}
    \end{overpic}
    \begin{overpic}[width=0.9\linewidth]{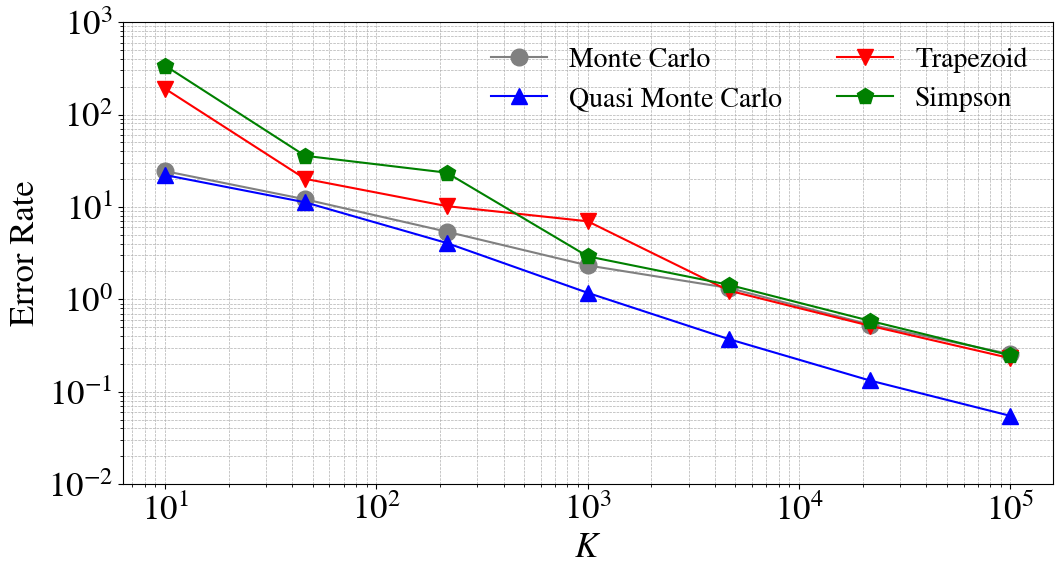}
        \put(1,55){\textbf{(b)}}
    \end{overpic}
    \caption{Error rate versus the number of the different Hadamard tests $K$ for each integration method including statistical error for estimating the ground state properties: (a) $M=1$ and (b) $M=1000$. $M$ is the number of shots of each Hadamard test.}
    \label{fig:ground_exact2}
\end{figure}

First, Fig.~\ref{fig:ground_exact} shows the results of numerical experiments focused solely on the numerical integration error.
The error rate is defined as the absolute difference between the estimated value and the desired value, divided by the desired value.
These results do not include statistical error, corresponding to the case where the number of Hadamard test measurements $M$ is infinite.
The vertical axis indicates the average error rate of the $100$ estimation trials, and the horizontal axis represents the number of variable samples $K$, or in other words the number of different Hadamard test circuits $K$.
Both the trapezoid and Simpson's rules exhibit a sharp drop in the error rate as the sample number $K$ increases, and eventually output highly accurate results.
It has been pointed out that this exponential convergence occurs when the integrand is an oscillatory and decaying function.
Since the Gaussian type discussed here has exactly such an integrand form, we can expect this phenomenon to be widely observed in applications involving the Fourier transform of Gaussian-type functions.
Additionally, the starting point of the convergence on the trapezoid rule is mostly the case where the integration width becomes smaller than the oscillation period, which is called as Nyquist limit~\cite{trefethen2014exponentially}.
Both MC and QMC converge steadily as $K$ increases, but QMC achieves a lower error rate more rapidly.

Next, we present numerical integration results that include statistical errors, as shown in Fig.~\ref{fig:ground_exact2}.
Panels (a) and (b) correspond to cases with $M=1$ and $M=1000$, respectively.
The case of $M=1$ is examined to observe the balance between statistical and numerical integration errors, while $M=1000$ serves as a representative example of a natural hardware setup.
In panel (a) ($M=1$), all estimation points show higher error rates compared to those in Fig.~\ref{fig:ground_exact}, indicating that statistical errors are dominant.
Here, MC and QMC exhibit lower errors, whereas the trapezoid and Simpson's rules display even larger errors.
Panel (b) presents the results for $M=1000$, where statistical errors become relatively small, allowing the numerical integration error to be observed. 
However, $K$ increases, the overall error eventually should become dominated by statistical error again except for MC, because the numerical errors converge faster and are asymptotically getting lower than the statistical errors.
In the end, QMC achieves the lowest overall in this setup.

\subsection{Case 2: Green's Function}
We next conduct a numerical experiment for estimating Green’s functions \cite{diag, keen2021quantumalgorithmsgroundstatepreparation}.
Since the Green’s function involves the inverse of the Hamiltonian, its evaluation benefits from employing a Fourier type integral representation of the inverse matrix.
This representation is not only applicable to Green’s functions but also extendable to more general inverse matrix computations, making this numerical experiment broadly significant.
More specifically, to approximate $H^{-1}$, we use the following truncated Fourier representation for an invertible Hermitian $H$:
\begin{align}
H^{-1} = -\,\frac{i}{\sqrt{2\pi}}\int_{0}^{y_{\rm c}} dy \int_{-z_{\rm c}}^{z_{\rm c}} dz \; z\, e^{-z^2/2}\, e^{-i y z H},    
\end{align}
This equation holds as $y_{\rm c},z_{\rm c}\to\infty$.
Here, the eigenvalues of $H$ are assumed to lie within $[-a,-1/b]\cup[1/b,a]$ with $y_{\rm c}=\Theta\!\big(b\sqrt{\log(b/\varepsilon)}\big)$ and $z_{\rm c}=\Theta\!\big(\sqrt{\log(b/\varepsilon)}\big)$ ensuring an $\varepsilon$-accurate truncation.
In our numerical experiments, we estimate the Green’s function
\begin{align}
G^{(+)}_{kl}(\omega) := \langle E_0 \,|\, \hat{a}_k \big(\omega-(H-E_0)+i\eta\big)^{-1} \hat{a}^{\dagger}_l \,|\, E_0\rangle,    
\end{align}
as discussed in Ref. \cite{diag}.
The Hamiltonian is the Heisenberg model of Eq.~(\ref{eq:heisenberg}), and the ground state $|E_0\rangle$ is computed analytically in advance.
We set the density matrix $\rho=|E_0\rangle\langle E_0|$.
For spins, $\hat a_k=(X_k+iY_k)/2$ and $\hat a^{\dagger}_l=(X_l-iY_l)/2$, but here we replace $\hat a_k,\hat a^{\dagger}_l$ by $X_k,X_l$ to isolate integration-scheme effects.
This does not affect our comparative conclusions.
As typical parameters, we take $k=l=0$, $\omega=E_0$, and $\eta=0.01$.
Hence we estimate $\mathrm{Tr}\!\big[X_0\,(2E_0+0.01\,i-H)^{-1} X_0\,|E_0\rangle\langle E_0|\big]$. 
In the form of Eq.~(\ref{eq:basic2}), with $V=[0,y_{\rm c}]\times[-z_{\rm c},z_{\rm c}]$ and $Z(z_{\rm c}):=1-e^{-z_{\rm c}^{2}/2}$, this is expressed as
\begin{align}
F(H) &= X_0\,(2E_0+0.01\,i-H)^{-1} X_0, \\
\label{eq:green_ft}
f(\bm t) &= \frac{1}{2\,y_{\rm c}\,Z(z_{\rm c})}\,|z|\,e^{-\frac{z^{2}}{2}}, \\
G(H,\bm t) &= -\,\frac{2\,i\,Z(z_{\rm c})}{\sqrt{2\pi}} \notag \\
&\ \ \ \ \ \times \mathrm{sign}(z)\,y_{\rm c}\; X_0\, e^{-i y z (2E_0+0.01\,i-H)}\, X_0,
\end{align}
for $\bm t=(y,z)$.
Since $G$ is the imaginary unit, the Hadamard test uses a circuit that extracts the imaginary value.
We set $y_{\rm c}=z_{\rm c}=8$, which was sufficient to make the truncation error negligible at double precision in our experiments.

\begin{figure}[t]
    \centering
    \includegraphics[width=0.9\linewidth]{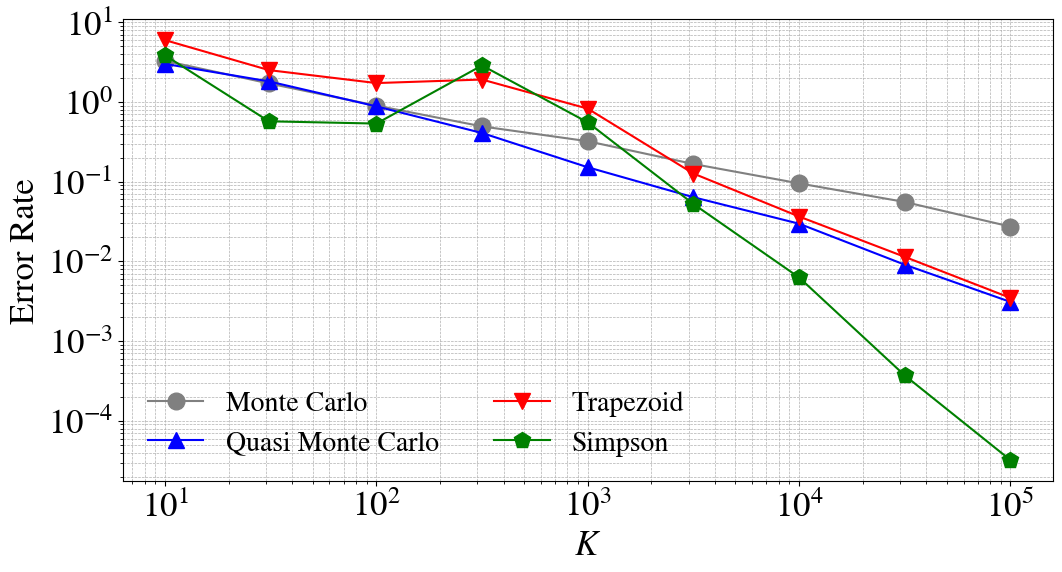}
    \caption{Error rate versus the number of the different Hadamard tests $K$ for each integration method without statistical error for estimating the Green's function. This corresponds to the limit case where $M \rightarrow \infty$.}
    \label{fig:green_exact}
\end{figure}

First, Fig.~\ref{fig:green_exact} presents the results of numerical experiments focusing solely on numerical integration error.
This situation corresponds to the case where $M$ is infinite.
In the trapezoid and Simpson's rules, the error rates decrease as $K$ increases; however, unlike the previous section, no exponential decay is observed.
This is because exponential decay occurs only when the integrand decays in each dimension~\cite{trefethen2014exponentially}, and whereas the function $f(\bm{t})$ in Eq. (\ref{eq:green_ft}) does not decay with respect to $y$.
Also, Simpson’s rule becomes advantageous as $K$ grows.
The converging begins from $K \geq 10^3$, which means the Nyquist limit blocks the convergence until $K < 10^3$.
Both MC and QMC steadily converge with increasing $K$, but QMC achieves a lower error rate more rapidly.

\begin{figure}[t]
    \centering
    \begin{overpic}[width=0.9\linewidth]{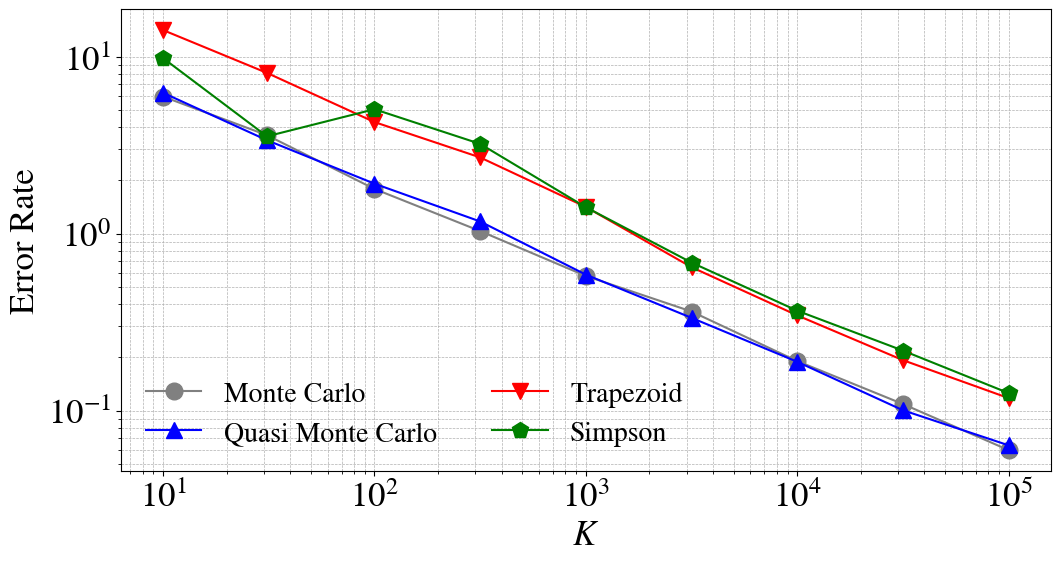}
        \put(1,55){\textbf{(a)}}
    \end{overpic}
    \begin{overpic}[width=0.9\linewidth]{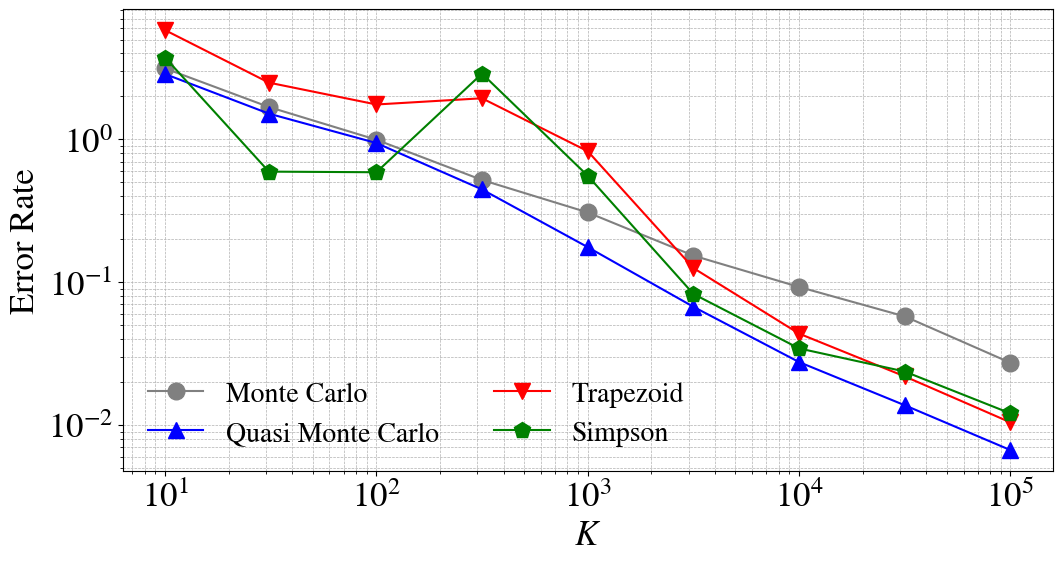}
        \put(1,55){\textbf{(b)}}
    \end{overpic}
    \caption{Error rate versus the number of the different Hadamard tests $K$ for each integration method including statistical error for estimating the Green's function: (a) $M=1$ and (b) $M=100$. $M$ is the number of shots of each Hadamard test.}
    \label{fig:green_m}
\end{figure}

Next, we present the results that also include statistical error, as shown in Fig.\ref{fig:green_m}.
Panel (a) of Fig.\ref{fig:green_m} ($M=1$) shows that the error rates at all estimation points are higher compared to Fig.\ref{fig:green_exact}.
Although the difference in error rates is not as significant as in the previous section, statistical error remains the dominant contribution.
Panel (b) of Fig.\ref{fig:ground_exact2} displays the results for $M=100$, where statistical error becomes relatively small and the impact of numerical integration error becomes apparent.
However, as in the previous section, increasing $K$ ultimately leads to overall error being dominated by the statistical error once again.
Consequently, QMC achieves the lowest error rate for large values of $K$.
As a result, QMC achieves the lowest overall error in this setup as well.

\subsection{Practical Considerations}
So far, we have seen that in the case of QMC, the error remains the smallest for $M \leq 1000$ in case 1, and for $M \leq 100$ in case 2. 
These applications employ Gaussian and inverse matrix Fourier transforms, which are commonly used in previous studies.
Therefore, similar discussions can be expected in other applications that utilize the same types of Fourier transforms.
Then, based on the results of our numerical experiments, we would like to further discuss the practical advantages of QMC from a more applied perspective.
In the case of the trapezoid rule, the Nyquist condition should be satisfied for the method to start converging.
However, in practice, it is often difficult to know the period of the integrand in advance.
In these applications, the shortest period of the integrand depends on the largest eigenvalue of the Hamiltonian $H$.
To satisfy the Nyquist condition, the integration width in each dimension should be as small as $|V|/K^{1/d}$ when dividing the domain evenly, which means that as the largest eigenvalue increases, it becomes more difficult for the trapezoid rule to achieve convergence.
On the other hand, QMC converges at least as fast as MC, or even faster, and its convergence does not depend on $K$ in the same way.
This allows us to obtain the estimate even for small values of $K$, and it is easy to stop the calculation early if necessary.
For these reasons, QMC can be used as a practical alternative to MC, offering greater flexibility and reliability in real setups.

We note that in Sec.~\ref{subseq:pre_lcucpp} we discussed previous studies on applications with dimensionality up to $d=4$.
The arguments presented there are expected to become even more relevant for larger values of $d$.
However, applications requiring substantially higher dimensional integration remain unclear at present, and identifying such applications and assessing their suitability is left for future work.

\section{CONCLUSION}
\label{seq:con}
In this study, we proposed QMC for LCU-CPP.
The LCU-CPP framework can realize a wide range of nonunitary operators, such as Gibbs states and inverse matrix, through integral representations, while requiring relatively few quantum hardware resources. 
We first analyzed the errors of the trapezoid rule, MC, and QMC.
We showed that the errors consist of numerical integration error and statistical error, and we derived upper bounds for these two contributions.
From this analysis, we argued that when the number of Hadamard test shots per unitary $M$ is in a moderate range, QMC has the potential to achieve the lowest error.
Since this conclusion was based only on upper bounds, and since the definition of the moderate range of $M$ was unclear, we performed numerical experiments.
Specifically, we studied ground state property estimation and Green’s function estimation.
We found that QMC indeed achieves the lowest error around $M=10^3$ for the former and around $M=10^2$ for the latter.
These values of $M$ are realistic for actual quantum devices, especially when circuit execution and communication latencies are taken into account.
Overall, our results demonstrate that QMC can serve as the most effective integration strategy within the LCU-CPP framework in practical experimental settings.

\section{ACKNOWLEDGMENTS}
This work is supported by MEXT Quantum Leap Flagship Program (MEXTQLEAP) Grant No.
JPMXS0120319794, and JST COI-NEXT program Grant
No. JPMJPF201.

\clearpage

\bibliographystyle{unsrt}

\onecolumngrid

\appendix

\section{Hadamard Test}
\label{ap:hd}

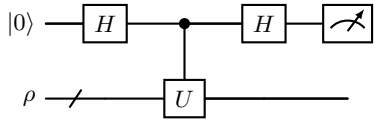
\begin{figure}
\centering
    \centering
    \begin{quantikz}
        \lstick{$\ket{0}$} & \gate{H}\qw & \ctrl{1} & \gate{H}\qw & \meter{} \\
        \lstick{$\rho$} & \qwbundle[]{} & \gate{U} & \qw & \qw
    \end{quantikz}
\caption{Quantum circuit for the Hadamard test. The outcome $s\in{+1,-1}$ satisfies $\Pr[s=+1]=(1+x)/2$ and $\Pr[s=-1]=(1-x)/2$, where $x = \mathrm{Re}[\mathrm{Tr}(U\rho)]$.}
\label{fig:hadamard_test2}
\end{figure}

We define a probability distribution $ F_x(s) $ parameterized by $ x $, where $ -1 \leq x \leq 1 $. The probability distribution is given by:
\begin{align}
\label{eq:ap_b_distribution}
F_x(s) = 
\begin{cases} 
    \frac{1-x}{2} & \text{for } s=-1 \\
    \frac{1+x}{2} & \text{for } s=+1
\end{cases} 
\end{align}
The expected value and variance of $ s $ under the distribution $ F_x(s) $ are as follows:
\begin{align}
\mathrm{Ex}[s] &= x, \\ 
\label{eq:ap_basic_var}
\mathrm{Var}[s] &= (1+x)(1-x)
\end{align}
where $ \mathrm{Ex}[s] $ denotes the expected value of $ s $ and $ \mathrm{Var}[s] $ denotes the variance of $ s $.
Considering $\rho$ as a quantum state and $ U $ as a unitary matrix, we estimate $\mathrm{Re}[U\rho] $.
An estimator $ \nu $ is defined by:
\begin{gather}
\nu = \frac{1}{K} \sum_{k=1}^{K} s_k
\end{gather}
where $ s_k $ is the $ k $-th outcome from the Hadamard test as depicted in Fig. $ \ref{fig:hadamard_test2} $, and it follows $ F_x(s_k) $.
The expected value and variance of $ \nu $ are as follows:
\begin{align}
\label{eq:ap_estimated}
\mathrm{Ex}[\nu] &= \frac{1}{K} \sum_{k=1}^{K} \mathrm{Ex}[s_k] = \mathrm{Re}[\mathrm{Tr}(U\rho)] , \\
\mathrm{Var}[\nu] &= \frac{1}{K} \sum_{k=1}^{K} \mathrm{Var}[s_k] = \frac{1}{K} \left( 1 - \mathrm{Re}[\mathrm{Tr}(U\rho)] ^2 \right).
\end{align}
The square root of this variance can be interpreted as a statistical error.

\section{Error Analysis in Monte Carlo Method}
\label{ap:mc}

The LCU-CPP approach relying on the Monte Carlo integration employs the following estimator:
\begin{align}
    \nu^{\rm mc} =\frac{1}{K} \sum_{k=1}^{K} \nu_k^{\rm mc},
\end{align}
where
\begin{gather}
\nu^{\rm mc}_k =  \frac{1}{M} \sum_{m=1}^{M} s_{m}(\bm{t}_k).
\label{eq:mc_estimator}
\end{gather}
Here, $s_{m}(\bm{t_k})\in \{1,-1\}$ is the output of the $m$-th execution of the Hadamard test for $\mathrm{Re}[G(A,\bm{t}_k)]$ where $\{\bm{t}_k\}$ represents mutually independent samples drawn from $f(\bm{t})$.
We seek to compute the expected value and the variance of $\nu^{\rm mc}$.

First, we consider the expected value.
The expected value of $s_{m}(\bm{t}_k)$ over the randomness of the measurement conditioned on the choice of $\bm{t}_k$ is:
\begin{align}
\mathrm{Ex}[s_{m}(\bm{t}_k)] &= \mathrm{Re}[\mathrm{Tr}(G(A,\bm{t}_k)\rho)] \label{eq:mc-expectation-value-over-measuremenet}
\end{align}
We can then take the expectation over the randomness of $\bm{t}_k$ for \eqref{eq:mc-expectation-value-over-measuremenet} to conclude $\mathrm{Ex}[s_{m}(\bm{t}_k)]=\mathrm{Tr}[F(A)\rho]$ and thus $\mathrm{Ex}[\nu_k^{\rm mc}] = \mathrm{Ex}[\nu^{\rm mc}]=\mathrm{Tr}[F(A)\rho]$.

Next, we look at the variance. 
Since $\nu_k^{\rm mc}$ are mutually independent and identically distributed, we have $\mathrm{Var}[\nu^{\rm mc}]=\mathrm{Var}[\nu^{\rm mc}_k]/K$.
It therefore suffices to calculate the variance of single $\nu^{\rm mc}_k$ to obtain $\mathrm{Var}[\nu^{\rm mc}]$.
Let us define a random variable $\tilde{\nu}^{\rm mc}:=\frac{1}{M}\sum_{m=1}^M s_m(\bm{t})$ where $\bm{t}$ is sampled from $f(\bm{t})$.
We make use of the law of total variance to compute $\tilde{\nu}^{\rm mc}$, which in this case can be written as:
\begin{align}
   \mathrm{Var}[\tilde{\nu}^{\rm mc}] = \mathrm{Ex}_{\bm{t}}[\mathrm{Var}_{\rm meas}(\tilde{\nu}^{\rm mc}(\bm{t}))] + \mathrm{Var}_{\bm{t}}[\mathrm{Ex}_{\rm meas}(\tilde{\nu}^{\rm mc}(\bm{t}))] \label{eq:law-of-total-variance}
\end{align}
$\mathrm{Var}_{\rm meas}[\tilde{\nu}^{\rm mc}(\bm{t})]$ can be calculated as
\begin{align}
    \mathrm{Var}_{\rm meas}[\tilde{\nu}^{\rm mc}(\bm{t})] = \frac{1}{M} \mathrm{Var}_{\rm meas}(s_{m}(\bm{t}))
\end{align}
since $s_m(\bm{t})$ represent independent execution of the Hadamard test and are thus independent random variables.
$\mathrm{Var}_{\rm meas}[s_{m}(\bm{t})]$ can be calculated as
\begin{align}
\begin{split}
    \mathrm{Var}_{\rm meas}[s_{m}(\bm{t})] &= \mathrm{Ex}[s_{m}^2(\bm{t})] - \mathrm{Ex}[s_{m}(\bm{t})]^2 = 1 - \mathrm{Re}[\mathrm{Tr}(G(A,\bm{t})\rho)]^2. 
\end{split}
\end{align}
We have
\begin{align}
    \mathrm{Ex}_{\bm{t}}[\mathrm{Var}_{\rm meas}(\tilde{\nu}^{\rm mc}(\bm{t}))] = \frac{1}{M} - \frac{1}{M}\int_V f(\bm{t}) \mathrm{Re}[\mathrm{Tr}(G(A,\bm{t})\rho)]^2 d\bm{t}. \label{eq:inner-variance}
\end{align}
Next, $\mathrm{Ex}_{\rm meas}(\tilde{\nu}^{\rm mc}(\bm{t}))$ can be calculated as,
\begin{align}
    \mathrm{Ex}_{\rm meas}(\tilde{\nu}^{\rm mc}(\bm{t})) = \frac{1}{M}\sum_{m=1}^M \mathrm{Ex}_{\rm meas}(s_m(\bm{t})) = \mathrm{Re}[\mathrm{Tr}(G(A,\bm{t})\rho)]
\end{align}
Therefore, 
\begin{align}
    \mathrm{Var}_{\bm{t}}[\mathrm{Ex}_{\rm meas}(\tilde{\nu}^{\rm mc}(\bm{t}))] &= \mathrm{Ex}_{\bm{t}}\left[\mathrm{Re}[\mathrm{Tr}(G(A,\bm{t})\rho)]^2\right] - \mathrm{Ex}_{\bm{t}}\left[\mathrm{Re}[\mathrm{Tr}(G(A,\bm{t})\rho)]\right]^2 \\
    &= \int_V f(\bm{t})\mathrm{Re}[\mathrm{Tr}(G(A,\bm{t})\rho)]^2 d\bm{t} - \mathrm{Tr}(F(A)\rho)^2 \label{eq:outer-variance}
\end{align}
From Eqs. \eqref{eq:law-of-total-variance}, \eqref{eq:inner-variance} and \eqref{eq:outer-variance}, we obtain
\begin{align}
    \mathrm{Var}[\tilde{\nu}^{\rm mc}] = \frac{1}{M} - \frac{1}{M}\int_V f(\bm{t}) \mathrm{Re}[\mathrm{Tr}(G(A,\bm{t})\rho)]^2 d\bm{t} + \int_V f(\bm{t})\mathrm{Re}[\mathrm{Tr}(G(A,\bm{t})\rho)]^2 d\bm{t} - \mathrm{Tr}(F(A)\rho)^2.
\end{align}
Finally, using that $\mathrm{Var}[\nu^{\rm mc}]=\mathrm{Var}[\nu^{\rm mc}_k]/K$,
\begin{align}
    \mathrm{Var}[\nu^{\rm mc}] = \frac{1}{K}\left(\int_V f(\bm{t})\mathrm{Re}[\mathrm{Tr}(G(A,\bm{t})\rho)]^2 d\bm{t} - \mathrm{Tr}(F(A)\rho)^2\right) + \frac{1}{KM}\left(1-\int_V f(\bm{t}) \mathrm{Re}[\mathrm{Tr}(G(A,\bm{t})\rho)]^2 d\bm{t}\right) 
\end{align}

\section{Error Analysis in Trapezoid Rule}
\label{ap:tr}
Firstly, we briefly review classical numerical integration error in the trapezoid rule.
Let $h : \mathbb{R}^d \rightarrow \mathbb{R}$ be a function with continuous second derivatives on a hyperrectangle $V = \prod_{j=1}^d [a_j, b_j]$.
We divide each axis into $N$ intervals of equal width $\Delta_j = (b_j - a_j)/(N-1)$, yielding $K = N^d$ grid points $\bm{x}_i$.
The trapezoid rule approximates the integral $I = \int_V h(\bm{x})\,d\bm{x}$ as
\begin{align}
T(K) = \Bigl(\prod_{j=1}^d \Delta_j \Bigr) \sum_{i=1}^K h(\bm{x}_i).
\end{align}
The error $E(K) := |I - T(K)|$ can be analyzed by Taylor expanding $h$ to second order around each grid point and integrating over each cell. This yields a local error per cell of
\begin{align}
E_\text{cell} = \sum_{j, j'=1}^d \frac{\partial^2 h}{\partial x_j \partial x_{j'}}(\bm{x}_i) \frac{\Delta_j \Delta_{j'}}{12} \prod_{m=1}^d \Delta_m + \mathcal{O}(\|\bm{\Delta}\|^4).
\end{align}
Summing over all $K$ cells gives
\begin{align}
E(K) \le \sum_{j,j'=1}^d \frac{|V|}{12} \max_{\bm{x}\in V} \left| \frac{\partial^2 h}{\partial x_j \partial x_{j'}} \right| \Delta_j \Delta_{j'} + \mathcal{O}(|\bm{\Delta}|^3),
\end{align}
where $|V| = \prod_{j=1}^d (b_j - a_j)$.
Using $\Delta_j \le 2(b_j - a_j)/K^{1/d}$, we have
\begin{align}
E(K) \le \frac{C}{K^{2/d}} + \mathcal{O}\left(\frac{1}{K^{3/d}}\right),
\end{align}
where
\begin{align}
C = \sum_{j,j'=1}^d \frac{|V|}{3} (b_j-a_j)(b_{j'}-a_{j'}) \max_{\bm{x}\in V} \left| \frac{\partial^2 h}{\partial x_j \partial x_{j'}} \right|,
\end{align}
which depends on the domain size and the maximum of the second derivatives of $h$.
Thus, the trapezoid rule converges as $\mathcal{O}(K^{-2/d})$.

Next, the LCU-CPP approach relying on the trapezoid rule employs the following estimator:
\begin{align}
\nu^{\rm tr} = \prod_{j=1}^d \Delta_j \sum_{k=1}^{K} \omega_k f(\bm{t}_k^{\rm{tr}}) \nu_k^{\rm tr},
\end{align}
where
\begin{gather}
\nu^{\rm tr}_k =  \frac{1}{M} \sum_{m=1}^{M} s_{m}(\bm{t_k}^{\rm tr}).
\end{gather}
Here, $\Delta_j = (b_j - a_j)/(K_j-1)$, where $K_j$ is the number of the integration points in the $j$-th dimension.
$\omega_k$ are the trapezoid rule weights for $\bm{t}_k^{\rm{tr}}$, and $\bm{t}_k^{\rm{tr}}$ is the $k$-th integration point.
$s_{m}(\bm{t_k}^{\rm tr})\in \{1,-1\}$ is the output of the $m$th execution of the Hadamard test for $G(A,\bm{t_k}^{\rm tr})$ where $\{\bm{t_k}^{\rm tr}\}$ represents evenly spaced grid points over $V$.

We then show the expected value.
The expected value of $s_{m}(\bm{t_k}^{\rm tr})$ over the deterministic integral point $\bm{t}_k$ is:
\begin{align}
\mathrm{Ex}[s_{m}(\bm{t_k}^{\rm tr})] &=  \mathrm{Re}[\mathrm{Tr}(G(A,\bm{t}_k)\rho)].
\label{eq:trap-expectation-value-over-measurement}
\end{align}
We obtain $\mathrm{Ex}[\nu_k^{\rm tr}] = \frac{1}{M} \sum_{m=1}^{M} \mathrm{Ex}[s_{m}(\bm{t_k}^{\rm tr})] = \mathrm{Re}[\mathrm{Tr}(G(A,\bm{t}_k)\rho)]$.
That leads to $\mathrm{Ex}[\nu^{\rm tr}] = \prod_{j=1}^d \Delta_j  \omega_k f(\bm{t}_k^{\rm{tr}}) \mathrm{Ex}[\nu_k^{\rm tr}] = \prod_{j=1}^d \Delta_j  \omega_k f(\bm{t}_k^{\rm{tr}}) \mathrm{Re}[\mathrm{Tr}(G(A,\bm{t}_k)\rho)]$, which is corresponding the trapezoid rule formula for estimating $\mathrm{Re}[\mathrm{Tr}(G(A,\bm{t}_k)\rho)]$.
Therefore, the expected value of $\nu_k^{\rm tr}$ is biased by the numerical integration error of the trapezoid rule.
Since the numerical integration error gives $\mathrm{Bias}^{\rm tr}(K) := \mathrm{Tr}[F(A)\rho] - \prod_{j=1}^d \Delta_j \sum_{k=1}^{K} \omega_k f(\bm{t}_k^{\rm{tr}})\mathrm{Re}[\mathrm{Tr}(G(A,\bm{t}_k)\rho)] = \mathrm{Tr}[F(A)\rho] - \mathrm{Ex}[\nu^{\rm tr}]$, the desired value $\mathrm{Tr}[F(A)\rho]$ can be written as:
\begin{align}
\mathrm{Tr}[F(A)\rho] = \mathrm{Ex}[\nu^{\rm tr}] + \mathrm{Bias}^{\rm tr}(K).
\end{align}
Therefore, we find the estimator is biased by the numerical integration error.
Next, we examine the variance of the estimator $\nu^{\rm tr}$. 
The variance is:
\begin{align}
\mathrm{Var}[\nu^{\rm tr}] = \left(\prod_{j=1}^d \Delta_j\right)^2 \sum_{k=1}^{K} \omega_k^2 f^2(\bm{t}_k^{\rm{tr}}) \mathrm{Var}[\nu_k^{\rm tr}] = \left(\prod_{j=1}^d \Delta_j\right)^2\sum_{k=1}^{K} \omega_k^2 f^2(\bm{t}_k^{\rm{tr}}) \frac{1}{M^2} \sum_{m=1}^{M} \mathrm{Var}[s_{m}(\bm{t_k}^{\rm tr})]. 
\end{align}
Since $\mathrm{Var}[s_{m}(\bm{t_k}^{\rm tr})] = \mathrm{Ex}[s_{m}(\bm{t_k}^{\rm tr})^2] - \mathrm{Ex}[s_{m}(\bm{t_k}^{\rm tr})]^2 = 1 - \mathrm{Re}[\mathrm{Tr}(G(A,\bm{t_k}^{\rm tr})\rho)]^2$, we can write:
\begin{align}
\mathrm{Var}[\nu^{\rm tr}] = \left(\prod_{j=1}^d \Delta_j\right)^2 \sum_{k=1}^{K} \omega_k^2 f^2(\bm{t_k}^{\rm tr}) \frac{1}{M} \left(1 - \mathrm{Re}[\mathrm{Tr}(G(A,\bm{t_k}^{\rm tr})\rho)]^2\right). 
\end{align}
Then, according to the numerical integration error,
\begin{align}
\mathrm{Var}[\nu^{\rm tr}] &= \prod_{j=1}^d \Delta_j \frac{1}{M} \int_{V} f^2(\bm{t}) \left(1 - \mathrm{Re}[\mathrm{Tr}(G(A,\bm{t})\rho)]^2\right) d\bm{t} + \mathcal{O}\left(\prod_{j=1}^d \Delta_j \frac{1}{MK^{\frac{2}{d}}}\right)
\end{align}
When all $K_j$ are assumed to be the same value, we can obtain $\prod_{j=1}^d \Delta_j = \prod_{j=1}^{d} (/b_j - a_j/ K_j-1) \leq 2^d \prod_{j=1}^{d} (b_j - a_j/K_j) = 2^d |V|/K$, and then we have 
\begin{align}
\mathrm{Var}[\nu^{\rm tr}] \leq \frac{1}{MK} 2^d |V| \int_{V} f^2(\bm{t}) \left(1 - \mathrm{Re}[\mathrm{Tr}(G(A,\bm{t})\rho)]^2\right) d\bm{t} + \mathcal{O}\left(\frac{1}{MK^{\frac{2}{d}+1}}\right).
 \end{align}
When the first term is dominant and we can see $\mathrm{Var}[\nu^{\rm tr}] = \mathcal{O}(1/MK)$. 
Considering the variance and the bias over the estimator, the mean squared error (MSE) to $\mathrm{Tr}[F(A)\rho]$ is:
\begin{align}
\mathrm{MSE}^{\rm tr} = \mathrm{Ex}[\left(\nu^{\rm tr} - \mathrm{Tr}[F(A)\rho] \right)^2] = \mathrm{Var}[\nu^{\rm tr}] + \mathrm{Bias}^{\rm tr}(K)^2,
\end{align}
where
\begin{align}
\mathrm{Bias}^{\rm tr}(K) \leq \frac{1}{K^{2/d}} \sum_{j, j'=1}^d
\frac{|V|(b_j -a_j)(b_{j'}-a_{j'})}{3}\,\max_{\bm{t} \in V}\left|\frac{\partial^2  f(\bm{t})\mathrm{Re}[\mathrm{Tr}(G(A,\bm{t})\rho)]}{\partial t_j\,\partial t_{j'}}\right| + \mathcal{O}\!\left(\frac{1}{K^{3/d}}\right).
\end{align}

\end{document}